\documentclass[conference,a4paper]{IEEEtran}

\usepackage[utf8]{inputenc} 
\usepackage[T1]{fontenc}
\usepackage{url}
\usepackage{ifthen}
\usepackage{cite}
\usepackage[cmex10]{amsmath} 
\usepackage{amssymb}

\newtheorem{thm}{Theorem}

\newtheorem{exmp}{Example}

\begin{document}
\title{An Optimal Linear Error Correcting Delivery Scheme for Coded Caching with Shared Caches} 

\author{%
  \IEEEauthorblockN{Nujoom Sageer Karat \IEEEauthorrefmark{1}, Spandan Dey\IEEEauthorrefmark{2}, Anoop Thomas \IEEEauthorrefmark{2} and B. Sundar Rajan \IEEEauthorrefmark{1}}
  \IEEEauthorblockA{\IEEEauthorrefmark{1}Department of Electrical Communication Engineering, Indian Institute of Science, Bengaluru 560012, KA, India \\
E-mail: \{nujoom,bsrajan\}@iisc.ac.in}

 \IEEEauthorblockA{\IEEEauthorrefmark{2} School of Electrical Sciences,  Indian Institute of Technology, Bhubaneswar 752050, OD, India \\
E-mail: \{anoopthomas,sd40\}@iitbbs.ac.in}
}

\maketitle

\begin{abstract}
Classical coded caching setting avails each user to have one dedicated cache. This is generalized to a more general shared cache scheme and the exact expression for the worst case rate was derived in [E. Parrinello, A. Unsal, P. Elia, ``
Fundamental Limits of Caching in Heterogeneous Networks with Uncoded Prefetching," available on  arXiv:1811.06247 [cs.IT], Nov. 2018]. For this case, an optimal linear error correcting delivery scheme is proposed and an expression for the peak rate is established for the same. Furthermore, a new delivery scheme is proposed, which gives an improved rate for the case when the demands are not distinct.  
\end{abstract}
\section{INTRODUCTION}
The technique of coded caching introduced in \cite{MaN} helps in reducing the peak traffic experienced by networks. This is achieved by making a part of
 the content locally available at the users during non-peak
 periods. In \cite{MaN}, it is shown that apart from the \emph{local caching gain} obtained by placing contents at user caches before the demands are revealed, a \emph{global caching gain} can be obtained by coded transmissions. The scheme in \cite{MaN} is a centralized coded caching scheme, where all users are linked to a single fixed server. Since then there have been many extensions to this, like decentralized scheme \cite{MaN2}, non-uniform demands \cite{NiM} and online coded caching \cite{PMN}.

A coded caching scheme involves two phases:  a placement phase and a delivery phase. In the placement phase or prefetching phase, each user can fill their local cache memory using the entire database. During this phase there is no bandwidth constraint as the network is not congested and the only constraint here is the memory. Delivery phase is carried out once the users reveal their demands. During the delivery phase only the server has access to the file database and the constraint here is the bandwidth as the network is congested in this phase. During placement phase some parts of files have to be judiciously cached at each user in such a way that the rate of transmission is reduced during the delivery phase. The prefetching can be done with or without coding. If during prefetching, no coding of parts of files is done, the prefetching scheme is referred to as uncoded prefetching \cite{MaN, YMA}. If coding is done during prefetching stage, then the prefetching scheme is referred to as coded prefetching \cite{CFL, JV, TC, ZT}.

An extension of the coded caching problem  involving
heterogeneous networks is considered in \cite{PUE}, where multiple users share a common cache. Each user has access to a helper cache, which is potentially accessed by multiple users. The scheme introduced in \cite{PUE} is referred to
as Shared Cache (SC) scheme throughout the paper. The corresponding prefetching scheme and delivery scheme are referred to as the SC prefetching scheme and SC delivery scheme respectively. In addition to the cache placement and delivery phase, there is an additional intermediate step which is the \emph{user-to-cache association phase}. The expression for rate in this scenario under the assumption of uncoded placement is derived in \cite{PUE}. The rate expression was under the assumption of worst case demand, which means that all the files are demanded. In our work, a new delivery scheme is
proposed for the non-distinct demand case which provides
improved rate compared to the SC scheme (Section \ref{sec:new_scheme}).

Error correcting coded caching scheme was introduced in \cite{KTR, KTR_2}. In this set up, the delivery phase is assumed to be error-prone and placement is assumed to be error-free. A similar model in which the delivery phase takes place over a packet erasure broadcast channel was considered in \cite{BWT}. In this work, shared cache systems in which the
delivery phase is error prone is considered.
An error correcting delivery scheme has to be designed to correct the required transmission errors. Each user has to decode their demands even in the presence of these errors. In our work, an optimal error correcting delivery scheme is proposed for the worst case demand in the shared cache system. 

The main contributions of this paper are as follows:
\begin{itemize}
	\item An optimal linear error correcting delivery scheme for coded caching problems with SC prefetching is proposed using techniques from index coding  (Section \ref{sec:alpha_sc} and Section \ref{sec: error_corr}).
	\item  For error correcting delivery scheme for coded caching problems with SC prefetching, a closed form expression for peak rate is established (Section \ref{sec: error_corr}).
	\item A new delivery scheme for SC prefetching for all the demand cases having an improved rate compared to the scheme in \cite{PUE} is proposed (Section \ref{sec:new_scheme}).
\end{itemize}

In this paper $\mathbb{F}_{q}$ denotes the finite field with $q$ elements, where $q$ is a power of a prime, and $\mathbb{F}^{*}_{q}$ denotes the  set of all non-zero elements of $\mathbb{F}_q$. 
For any integer $K$, let $[K]$ denote
the set $\{1,2, \ldots,K\}.$
For a $K \times N$ matrix $L$, $L_i$ denotes its $i$th row. Also, ${ n \choose k} \triangleq \frac{n!}{(n-k)!k!}$ and ${n \choose k} =0 $ if $n <k$. The lower convex envelope of points $\{(i, f(i)): i \in [n] \cup \{0\}\}$ for some natural number $n$ is denoted by $Conv(f(i))$.

A linear $[n,k,d]_q$ code $\mathcal{C}$ over $\mathbb{F}_q$ is a $k$-dimensional subspace of $\mathbb{F}^{n}_{q}$ with minimum Hamming distance $d$. A matrix ${G}$ of size $k \times n$ whose rows are linearly independent codewords of $\mathcal{C}$ is called a generator matrix of $\mathcal{C}$. A linear $[n,k,d]_q$ code $\mathcal{C}$ can thus be represented using its generator matrix ${G}$ as,
$ \mathcal{C} = \{ {\mathbf{y}}G: {\mathbf{y}} \in \mathbb{F}^{k}_{q} \}.$ 
Let $N_{q}[k,d]$ denote the length of the shortest linear code over $\mathbb{F}_q$ which has dimension $k$ and minimum distance $d$. 

%%%%%%%%%%%%%%%%%%%%%%%%%%%%%%%%%%%%%%%%%%%%%%%%%%%%%%%%%%%%%%%%%%%%%
\section{Preliminaries and Background}
\label{sec:prelim}
To obtain the main results of this paper, we use results from error correcting index coding problems \cite{DSC}. In this section we recall some results from this and also review the concepts of error correcting coded caching scheme \cite{KTR}. Furthermore, we review the SC placement and delivery scheme \cite{PUE}.
%%%%%%%%%%%%%%%%%%%%%%%
\subsection{Index Coding Problem}
The index coding problem with side information was introduced in \cite{BiK}. A single source has $n$ messages $x_1,x_2 \ldots , x_n$ where $x_i \in \mathbb{F}_{q}, ~\forall i \in [n].$ There are $K$ receivers, $R_1, R_2, \ldots, R_K$. Each receiver possesses a subset of messages as side information. Let $\mathcal{X}_i$ denote the set of  indices of the messages belonging to the side information of receiver $R_i$. The map $f:[K] \rightarrow [n] $ assigns receivers to indices of messages demanded by them. Receiver $R_i$ demands the message $x_{f(i)}$, $f(i) \notin \mathcal{X}_i$ \cite{DSC}. The source knows the side information available to each receiver and has to satisfy the demand of each receiver in minimum number of transmissions. An instance of index coding problem can be completely characterized by a side information hypergraph \cite{AHLSW}. Given an instance of the index coding problem, finding the best \textit{scalar linear} binary index code is equivalent to finding the \textit{min-rank} of the side information hypergraph \cite{DSC}, which is known to be an NP-hard problem in general \cite{BBJ, RP, DSC2}.

%The index coding problem with side information was introduced in \cite{BiK}. A  source has to broadcast some messages to a set of receivers. Each receiver has some messages as side information. The source knows the side information available to each receiver and has to satisfy the demand of each receiver in minimum number of transmissions. 
%
%In the case of index coding problem, the message vector $X \in \mathbb{F}_q^{K \times 1}$ is possessed by a single source. There are $m$ receivers and $K$ messages. The $i$th receiver $R_i$ possesses the messages indexed by $\mathcal{X}_i$. The map $f:[m] \rightarrow [K] $ assigns users to indices of messages demanded by them. This means that user $i$ demands the message $X_{f(i)}$, $f(i) \notin \mathcal{X}_i$ \cite{DSC}.

An index coding problem with $K$ receivers and $n$ messages can be represented by a hypergraph $\mathcal{H} (V,E)$, where $V=[n]$ is the set of vertices and $E$ is the set of hyperedges \cite{AHLSW}. Vertex $i$ represents the message $x_i$ and each hyperedge represents a receiver. In \cite{DSC}, the min-rank of a hypergraph $\mathcal{H}$ over $\mathbb{F}_q$ is defined as,
\begin{equation*}
\kappa(\mathcal{H}) \triangleq
%\min\{\text{rank}_{q}(\{\mathbf{v_i}\}) \}
\min\{\text{rank}_q(\{{\mathbf{v_i}}
+{\mathbf{e_{f(i)}}}\}_{i\in [K]}): \\
{\mathbf{v_i}} \in \mathbb{F}^{n}_q, {\mathbf{v_i}} \triangleleft \mathcal{X}_i\},
\end{equation*}
where $\bf{v_i}$ $\triangleleft$ $\mathcal{X}_i$ denotes that $\bf{v_i}$ is the subset of the  support of $\mathcal{X}_i$; the support of a vector $\bf{u}$ $\in \mathbb{F}^{n}_{q}$ is defined to be the set $\{i\in [n]: u_i \neq 0  \}$. This min-rank defined above is the smallest length of scalar linear index code for the problem. A linear index code of length $N$ can be expressed as $XL$, where $L$ is an $n \times N$ matrix and $X = [x_1~x_2 \ldots ~x_n]$. The matrix $L$ is said to be the \textit{matrix corresponding to the index code}. 

Let $\mathcal{G}$ = $(\mathcal{V},\mathcal{E})$ be an undirected graph, then a subset of vertices $\mathcal{S}$ $\subseteq$ $\mathcal{V}$ is called an independent set if $\forall u, v \in \mathcal{S}$, $\{u,v\}$ $\notin$ $\mathcal{E}$. The size of a largest independent set in the graph $\mathcal{G}$ is called the independence number of $\mathcal{G}$. Dau {\it{et al}}. in \cite{DSC} extended the notion of independence number to the case of directed hypergraph corresponding to an index coding problem. For each receiver $R_i$, define the sets $$
\mathcal{Y}_i \triangleq [n] \setminus \bigg( \{f(i) \} \cup \mathcal{X}_i \bigg) $$
and
$$\mathcal{J(\mathcal{H})} \triangleq \cup_{i\in [K]} \{\{f(i)\} \cup Y_{i} : Y_i \subseteq \mathcal{Y}_i\}.$$
A subset $H$ of $[n]$ is called a generalized independent set in $\mathcal{H}$, if every nonempty subset of $H$ belongs to $\mathcal{J(\mathcal{H})}$. The size of the largest independent set in $\mathcal{H}$ is called the generalized independence number and is denoted by $\alpha (\mathcal{H})$. It is proved in \cite{KTR} that for any index coding problem,
\begin{equation}
\label{eq:alphaleqkappa}
\alpha (\mathcal{H}) \leq \kappa (\mathcal{H}).
\end{equation}

The quantities  $\alpha (\mathcal{H})$ and $\kappa(\mathcal{H})$ decide the bounds on the optimal length of error correcting index codes. The error correcting index coding problem  with side information was defined in \cite{DSC}. An index code is said to correct $\delta$ errors if after receiving at most $\delta$ transmissions in error, each receiver is able to decode its demand. A $\delta$-error correcting index code is represented as $(\delta, \mathcal{H})$-ECIC. An optimal linear $(\delta, \mathcal{H})$-ECIC over $\mathbb{F}_q$ is a linear $(\delta, \mathcal{H})$-ECIC over $\mathbb{F}_q$ of the smallest possible length $\mathcal{N}_{q}[\mathcal{H},\delta]$. 
Lower and upper bounds on  $\mathcal{N}_{q}[\mathcal{H},\delta]$ were established in \cite{DSC}. The Lower bound is known as the $\alpha$-bound and the upper bound is known as the $\kappa$-bound.  
The length of an optimal linear $(\delta,\mathcal{H})$-ECIC over $\mathbb{F}_q$ satisfies
\begin{equation}
\underbrace{N_q[\alpha(\mathcal{H}), 2\delta + 1]~ \leq ~}_{\alpha\text{-bound}}  \mathcal{N}_{q}[\mathcal{H},\delta] \underbrace{~\leq~ N_q[\kappa(\mathcal{H}), 2\delta + 1]}_{\kappa\text{-bound}}. \label{eq:bds}
\end{equation}
The $\kappa$-bound is achieved by concatenating an optimal linear classical error correcting code and an optimal linear index code. Thus for any index coding problem, if $\alpha (\mathcal{H})$ is same as $\kappa_q(\mathcal{H})$, then concatenation scheme would give optimal error correcting index codes \cite{SaR, SagR, SSR, KSR}.

%%%%%%%%%%%%%%%%%%%%%%%
\subsection{Error Correcting Coded Caching Scheme}
Error correcting coded caching scheme was proposed in \cite{KTR}. The server is connected to $K$ users through a shared link which is error prone. 
The server has access to $N$ files $X^1, X^2, \ldots, X^N$, each of size $F$ bits. Every user has an isolated cache with memory $MF$ bits, where $M \in [0,N]$.  A prefetching scheme is denoted by ${\mathcal{M}}$. During the delivery phase, only the server has access to the database. Every user demands one of the $N$ files. The demand vector is denoted by $\mathbf{d} = (d_1, \ldots, d_K)$, where $d_i$ is the index of the file demanded by user $i$. The number of distinct files requested in $\mathbf{d}$ is denoted by $N_e(\mathbf{d})$. During the delivery phase, the server informed of the demand $\mathbf{d}$, transmits a function of $X^1, \ldots, X^N$, over a shared link. Using the cache contents and the transmitted data, each user $i$ needs to reconstruct the requested file $X^{d_{i}}$ even if $\delta$ transmissions are in error.

For the $\delta$-error correcting coded caching problem, a communication rate $R(\delta)$ is \textit{achievable} for demand $\mathbf{d}$ if and only if there exists a transmission of $R(\delta)F$ bits such that every user $i$ is able to recover its desired file $X^{d_{i}}$ even after at most $\delta$ transmissions are in error. Rate  $R^*(\mathbf{d}, \mathcal{M}, \delta)$ is the minimum achievable rate for a given $\mathbf{d}$, $\mathcal{M}$ and $\delta$. The average rate $R^*(\mathcal{M}, \delta)$ is defined as the expected minimum average rate given $\mathcal{M}$ and $\delta$ under uniformly random demand. Thus $ R^*(\mathcal{M}, \delta) = \mathbb{E}_{\mathbf{d}}[R^*(\mathbf{d}, \mathcal{M}, \delta)].$

The average rate depends on the prefetching scheme $\mathcal{M}$. The minimum average rate  $ R^*(\delta)= \min_{\mathcal{M}} R^*(\mathcal{M}, \delta)$  is the minimum rate of the delivery scheme over all possible $\mathcal{M}$. The rate-memory trade-off for average rate is finding the minimum average rate $R^*(\delta)$ for different memory constraints $M$. Another quantity of interest is the peak rate, denoted by $R^*_{\text{worst}}(\mathcal{M}, \delta)$, which is defined as
$R^*_{\text{worst}}(\mathcal{M}, \delta) = \max_{\mathbf{d}} R^*(\mathbf{d}, \mathcal{M}, \delta).$ 
The minimum peak rate is defined as
$ R^*_{\text{worst}}(\delta)= \min_{\mathcal{M}} R^*_{\text{worst}}(\mathcal{M}, \delta).$

%%%%%%%%%%%%%%%%%%%%%%%
\subsection{Shared Cache Scheme}
The coded caching system with shared cache \cite{PUE} is described as follows. There are $N$ files, $K$ users and $\Lambda \leq K$ caches, with normalized memory of each cache being $M$. Parameter $\gamma$ is defined to be $\gamma = \frac{M}{N}$. 
Each cache $\lambda =1,2, \ldots, \Lambda$, is assigned to a set of users $\mathcal{U}_\lambda$, and all these disjoint sets,
$$ \mathcal{U} \triangleq \{\mathcal{U}_1, \mathcal{U}_2, \ldots, \mathcal{U}_\Lambda \} $$
form a partition of the set of users $\{1,2, \ldots, K\},$ describing the overall association of the users to the caches. For any given $\mathcal{U}$, we consider the association profile 
$$ \mathcal{L} =(\mathcal{L}_1, \ldots, \mathcal{L}_{\Lambda})$$ 
where $\mathcal{L}_\lambda$ is the number of users assigned to the $\lambda$th most populated helper node/cache.

1) \emph{SC Prefetching Phase}: Each file $X^n$ is split into ${\Lambda \choose \Lambda \gamma}$ disjoint subfiles $X^n_{\mathcal{T}}$, for each $\mathcal{T} \subset [\Lambda]$, $|\mathcal{T}|=\Lambda \gamma$, and then each cache stores a fraction $\gamma$ of each file. For instance, the $\lambda$th cache stores subfiles in the set $\{X^n_{\mathcal{T}}: \lambda \in \mathcal{T}, \forall n \in [N]\}.$  This prefetching scheme is denoted by $\mathcal{M}_{\text{SC}}$. 

2) \emph{SC Delivery Phase}: Without loss of generality assume $|\mathcal{U}_1| \geq |\mathcal{U}_2| \geq \ldots |\mathcal{U}_\Lambda|$ (any  other case can be handled by simple relabeling of the caches) and  $\mathcal{L}_\lambda = |\mathcal{U}_\lambda|$. With a slight abuse of notation, each $\mathcal{U}_\lambda$ denotes an ordered vector describing the users associated to cache $\lambda$. Delivery phase consists of $\mathcal{L}_1$ rounds, where each round $j \in [\mathcal{L}_1]$ serves users 
$$ \mathcal{R}_j= \bigcup_{\lambda \in [\Lambda]} (\mathcal{U}_\lambda (j): \mathcal{L}_\lambda \geq j), $$
where $\mathcal{U}_\lambda (j)$ is the $j$th user in the set $\mathcal{U}_\lambda.$ For each round $j$, the sets $\mathcal{Q} \subseteq [\Lambda]$ of size $|\mathcal{Q}| = \Lambda \gamma +1$ are considered and for each set $\mathcal{Q}$ . The set of receiving users are
$$ \mathcal{E}_\mathcal{Q} = \bigcup_{\lambda \in \mathcal{Q}} (\mathcal{U}_\lambda (j): \mathcal{L}_\lambda \geq j). $$
If  $\mathcal{E}_\mathcal{Q} \neq \phi$, the server transmits,
$$ T_{\mathcal{E}_\mathcal{Q}} = \oplus_{\lambda \in \mathcal{Q}: {\mathcal{L}_\lambda} \geq j} X^{d_{\mathcal{U}_\lambda(j)}}_{\mathcal{Q}\setminus \{\lambda\}}. $$
If $\mathcal{E}_\mathcal{Q} = \phi$, there is no transmission. 
The decoding is possible for each user using these transmissions \cite{PUE}. 
The optimal worst case rate for the SC scheme is obtained in \cite{PUE} as
$$R^*_{\text{worst}}(\mathcal{M}_{\text{SC}}, 0) = Conv\bigg( \frac{\sum_{i=1}^{\Lambda-\Lambda \gamma} \mathcal{L}_i {\Lambda-i \choose \Lambda \gamma}}{{\Lambda \choose \Lambda \gamma}}\bigg)$$
at points $\gamma =\{\frac{1}{\Lambda}, \frac{2}{\Lambda}, \ldots, 1\}.$

For a fixed prefetching $\mathcal{M}$ and for a fixed demand $\mathbf{d}$, the delivery phase of a coded caching problem is an index coding problem \cite{MaN}. In fact, for fixed prefetching, a coded caching scheme consists of $N^K$  parallel index coding problems one for each of the $N^K$ possible user demands. Thus finding the minimum achievable rate for a given demand $\mathbf{d}$ is equivalent to finding the min-rank of the equivalent index coding problem induced by the demand $\mathbf{d}$.

Consider the SC prefetching scheme $\mathcal{M}_{\text{SC}}$. The index coding problem induced by the demand $\mathbf{d}$  for SC prefetching is denoted by $\mathcal{I}(\mathcal{M}_{\text{SC}}, \mathbf{d}).$ Each subfile $X^n_{\mathcal{T}}$ corresponds to a message in the index coding problem. The corresponding generalized independence number and min-rank are represented as $\alpha(\mathcal{M}_{\text{SC}}, \mathbf{d})$ and $\kappa(\mathcal{M}_{\text{SC}}, \mathbf{d})$ respectively.

%%%%%%%%%%%%%%%%%%%%%%%%%%%%%%%%%%%%%%%%%%%%%%%%%%%%%%%%%%%%%%%%%%%%%
\section{Generalized Independence Number for $\mathcal{I}(\mathcal{M}_{\text{SC}}, \mathbf{d})$} 
\label{sec:alpha_sc}
In this section we find a closed form expression for generalized independence number $\alpha(\mathcal{M}_{\text{SC}}, \mathbf{d})$ of the index coding problem $\mathcal{I}(\mathcal{M}_{\text{SC}}, \mathbf{d})$ for the case when all the files are demanded. We denote the worst case demand vector as $\mathbf{d}_{\text{worst}}.$ Hence our aim is to find an expression for $\alpha(\mathcal{M}_{\text{SC}}, \mathbf{d}_{\text{worst}})$. In  $\mathcal{I}(\mathcal{M}_{\text{SC}}, \mathbf{d})$ each subfile corresponds to a message. The side information sets of all the receivers in the index coding problem is completely decided by the placement scheme in \cite{PUE}. We assume a unicast index coding problem for convenience (if there is a receiver demanding multiple messages, we split that receiver into multiple receivers each demanding one file). Hence there are $N {\Lambda \choose \Lambda \gamma}$ messages and $K {\Lambda \choose \Lambda \gamma}$ receivers in $\mathcal{I}(\mathcal{M}_{\text{SC}}, \mathbf{d})$. From the delivery scheme and the expression for rate in \cite{PUE}, we get an upper  bound for $\kappa(\mathcal{M}_{\text{SC}}, \mathbf{d}_{\text{worst}})$ as
\begin{equation}
\kappa(\mathcal{M}_{\text{SC}}, \mathbf{d}_{\text{worst}}) \leq \sum_{i=1}^{\Lambda-\Lambda \gamma} \mathcal{L}_i {\Lambda-i \choose \Lambda \gamma}.
\label{eq:kappa_sc}
\end{equation}

In the proof of the theorem below, we give a technique to find a generalized independent set for $\mathcal{I}(\mathcal{M}_{\text{SC}}, \mathbf{d}_{\text{worst}})$ by intelligently picking messages to the set. 
Using this we get a lower bound for the generalized independence number, $\alpha(\mathcal{M}_{\text{SC}}, \mathbf{d}_{\text{worst}}).$ From this we conclude that $\alpha(\mathcal{M}_{\text{SC}}, \mathbf{d}_{\text{worst}}) = \kappa(\mathcal{M}_{\text{SC}}, \mathbf{d}_{\text{worst}}).$
\begin{thm}
\label{thm:alphakappascw}
For the index coding problems $\mathcal{I}(\mathcal{M}_{\text{SC}}, \mathbf{d}_{\text{worst}})$ for the case when all the files are demanded, we have
$$\alpha(\mathcal{M}_{\text{SC}}, \mathbf{d}_{\text{worst}}) = \kappa(\mathcal{M}_{\text{SC}}, \mathbf{d}_{\text{worst}})=\sum_{i=1}^{\Lambda-\Lambda \gamma} \mathcal{L}_i {\Lambda-i \choose \Lambda \gamma}. $$
\end{thm}
\begin{IEEEproof}
We construct a set $B(\mathbf{d}_{\text{worst}})$ whose elements are messages of $\mathcal{I}(\mathcal{M}_{\text{SC}}, \mathbf{d}_{\text{worst}})$ such that the set of indices of the messages in $B(\mathbf{d}_{\text{worst}})$ forms a generalized independent set. The set $B(\mathbf{d}_{\text{worst}})$ is constructed as
\begin{align*}
  B(\mathbf{d}_{\text{worst}})&= \\ &\bigcup_{ i \in [N]} \{X^i_{\{a_1, \ldots, a_{\Lambda \gamma}\}}: a_1, \ldots, a_{\Lambda \gamma} \neq 1, 2, \ldots,c(i) \},
 \end{align*}
where $c(i)$ represents the cache to which the user demanding the file $X^i$ is associated with. For instance, if $X^i$ is connected to the $n-$th cache, then $c(i)=n$.
	Let $H(\mathbf{d}_{\text{worst}})$ be the set of indices of the messages in $B(\mathbf{d}_{\text{worst}})$. The claim is that  $H(\mathbf{d}_{\text{worst}})$ is a generalized independent set. Each message in $B(\mathbf{d}_{\text{worst}})$ is demanded by one receiver. Hence all the subsets of $H(\mathbf{d}_{\text{worst}})$ of size one are present in $\mathcal{J}(\mathcal{I})$.
	Consider any set $B' = \{X^{i_1}_{ \{a_{1_1}, \ldots, a_{1_{\Lambda \gamma}}\}}, \ldots, X^{i_k}_{\{a_{k_1}, \ldots, a_{k_{\Lambda \gamma}}\}}\}$ $\subseteq B(\mathbf{d}_{\text{worst}})$ where $i_1 \leq {i_2} \leq \ldots \leq {i_k}$. Consider the message $X^{i_1}_{\{a_{1_1}, \ldots, a_{1_{\Lambda \gamma}}\}}$. The receiver demanding this message does not have any other message in $B'$ as side information. Thus indices of messages in $B'$ lie in $\mathcal{J}(\mathcal{I})$. Thus any subset of $H(\mathbf{d}_{\text{worst}})$ lies in $\mathcal{J}(\mathcal{I})$. 	
	Since $H(\mathbf{d}_{\text{worst}})$ is a generalized independent set, we have, $\alpha({\mathcal{M}_{\text{SC}}, \mathbf{d}_{\text{worst}}}) \geq |H(\mathbf{d}_{\text{worst}})| $. Note that $|H(\mathbf{d}_{\text{worst}})|=|B(\mathbf{d}_{\text{worst}})|$. 
	
	Number of messages of the form $X^n_{\{a_1, ..., a_{\Lambda \gamma}\}}$ which are present in $B(\mathbf{d}_{\text{worst}})$ is ${\Lambda-c(n) \choose \Lambda \gamma}$. Hence,  of the $\mathcal{L}_i$ files demanded by the users which are associated to the $i$th cache, the number of subfiles or equivalently messages which are picked to the set $B(\mathbf{d}_{\text{worst}})$ is  $\mathcal{L}_i{\Lambda-i \choose \Lambda \gamma}$. Since ${\Lambda-i \choose \Lambda \gamma}$ is defined to be zero if $\Lambda -i \leq \Lambda \gamma$, the limits of summation only needs to be taken from $i=1$ to $\Lambda-\Lambda \gamma$.	
	Thus $$|B(\mathbf{d}_{\text{worst}})| = \sum_{i=1}^{\Lambda-\Lambda \gamma} \mathcal{L}_i{\Lambda-i \choose \Lambda \gamma}.$$
	Hence, $\alpha({\mathcal{M}_{\text{SC}}, \mathbf{d}_{\text{worst}}}) \geq \mathcal{L}_i{\Lambda-i \choose \Lambda \gamma}. $
	Hence from  \eqref{eq:alphaleqkappa} and \eqref{eq:kappa_sc}, the statement of the theorem follows.
\end{IEEEproof}

\begin{exmp}
\label{Ex:alphasc}
Consider a scenario with $K=N=8$, $M=\Lambda=4$ and $\mathbf{\mathcal{L}}=(3,2,2,1)$. In the placement phase, each file $X^n$ is first split into ${\Lambda \choose \Lambda \gamma}=6$ equally-sized subfiles\footnote{For simplicity we use $X^n_{1,2}$ instead of $X^n_{\{1,2\}}$.}: $ X^n_{1,2} , X^n_{1,3}, X^n_{1,4}, X^n_{2,3}, X^n_{2,4}, X^n_{3,4} $ and then each cache $\lambda$ stores $X^n_{\mathcal{T}}: \lambda \in \mathcal{T}, \forall n \in [8]$. For example, cache 1 stores subfiles $X^n_{1,2}, X^n_{1,3}, X^n_{1,4}$. In the cache assignment, users $\mathcal{U}_1=\{1,2,3\}, \mathcal{U}_2=\{4,5\}, \mathcal{U}_3=\{6,7\}$ and $\mathcal{U}_4=\{8\}$ are assigned to caches $1,2,3$ and $4$ respectively, so that the association profile is $\mathbf{\mathcal{L}}=(3,2,2,1)$. Without loss of generality we assume that the demand vector $\mathbf{d}_{\text{worst}}=(1,2,3,4,5,6,7,8)$. 

We consider the index coding problem $\mathcal{I}(\mathcal{M}_{\text{SC}}, \mathbf{d}_{\text{worst}})$. Each of the subfiles correspond to a message in the index coding problem. Hence for this example, the corresponding $\mathcal{I}(\mathcal{M}_{\text{SC}}, \mathbf{d}_{\text{worst}})$ will have 48 messages and 48 receivers (each user demanding more than one message is split into multiple receivers demanding one message each). We construct a set $B(\mathbf{d}_{\text{worst}})$, whose elements are messages of $\mathcal{I}(\mathcal{M}_{\text{SC}}, \mathbf{d}_{\text{worst}})$ such that the set of indices of the messages in $B(\mathbf{d}_{\text{worst}})$ forms a generalized independent set. The set $B(\mathbf{d}_{\text{worst}})$ for this case can be constructed as 
\begin{align*}
 B(\mathbf{d}_{\text{worst}})= &\{X^1_{2,3}, X^1_{2,4}, X^1_{3,4}, X^2_{2,3}, X^2_{2,4}, X^2_{3,4}, \\ & X^3_{2,3}, X^3_{2,4}, X^3_{3,4}, X^4_{3,4}, X^5_{3,4} \}.
\end{align*}  
Hence $\alpha(\mathcal{M}_{\text{SC}},\mathbf{d}_{\text{worst}}) \geq 11.$ From the transmission scheme in \cite{PUE}, there are 11 transmissions which satisfy the demands of all the users. Hence $\kappa(\mathcal{M}_{\text{SC}},\mathbf{d}_{\text{worst}}) \leq 11.$ Thus from \eqref{eq:alphaleqkappa} we have for this case, $\alpha(\mathcal{M}_{\text{SC}},\mathbf{d}_{\text{worst}}) = \kappa(\mathcal{M}_{\text{SC}},\mathbf{d}_{\text{worst}}) = 11.$   
\end{exmp}
\section{Optimal Error Correcting Delivery Scheme for SC Prefetching for Worst Case Demand}
\label{sec: error_corr}
For the worst case demand, we have proved in Theorem \ref{thm:alphakappascw} that $\alpha(\mathcal{M}_{\text{SC}},\mathbf{d}_{\text{worst}}) = \kappa(\mathcal{M}_{\text{SC}},\mathbf{d}_{\text{worst}})$. Hence for this case, the optimal linear error correcting delivery scheme can be constructed by concatenating the worst case delivery scheme in \cite{PUE} with an optimal error correcting code which corrects the required number of errors. Based on this we give an expression for the worst case rate for SC prefetching in the theorem below.
\begin{thm}
For a shared cache system with SC prefetching scheme, we have
$$R^*_{\text{worst}}(\mathcal{M}_{\text{SC}}, \delta) = Conv\bigg( \frac{N_q[\sum_{i=1}^{\Lambda-\Lambda \gamma} \mathcal{L}_i {\Lambda-i \choose \Lambda \gamma}, 2\delta+1]}{{\Lambda \choose \Lambda \gamma}}\bigg)$$
at points $\gamma =\{\frac{1}{\Lambda}, \frac{2}{\Lambda}, \ldots, 1\}.$
\end{thm}
\begin{IEEEproof}
From Theorem \ref{thm:alphakappascw}, we get that for $\mathcal{I}(\mathcal{M}_{\text{SC}}, \mathbf{d}_{\text{worst}}),$  $\alpha(\mathcal{M}_{\text{SC}}, \mathbf{d}_{\text{worst}})= \kappa(\mathcal{M}_{\text{SC}}, \mathbf{d}_{\text{worst}}).$   Thus from (\ref{eq:bds}), the $\alpha$ and $\kappa$ bounds become equal for such index coding problems. The optimal length or equivalently the optimal number of transmissions required for $\delta$ error corrections in those index coding problems is thus $N_q[\kappa(\mathcal{M}_{\text{SC}}, \mathbf{d}_{\text{worst}}), 2\delta+1]=N_q[\sum_{i=1}^{\Lambda-\Lambda \gamma} \mathcal{L}_i {\Lambda-i \choose \Lambda \gamma}, 2\delta+1]$  and hence the statement of the theorem follows.
\end{IEEEproof}

Since $\alpha$ and $\kappa$ bounds meet for $\mathcal{I}(\mathcal{M}_{\text{SC}}, \mathbf{d}_{\text{worst}}),$ the optimal linear error correcting delivery scheme here would be concatenation of SC delivery scheme with an optimal classical error correcting delivery scheme which corrects $\delta$ errors. Decoding can be done by syndrome decoding for error correcting index codes proposed in \cite{DSC}.
We give an example for which we construct optimal error correcting delivery scheme for coded caching problems with SC prefetching.

\begin{exmp}
Consider the coded caching problem with shared caches which we considered in Example \ref{Ex:alphasc}. For this we know that the $\alpha$ and $\kappa$ bounds meet and hence the concatenation scheme is optimal. For this case, the SC delivery scheme is as follows. There are 3 rounds with each round serving the following sets of users: $\mathcal{R}_1=\{1,4,6,8 \}, \mathcal{R}_2=\{2,5,7\}, \mathcal{R}_3=\{3\}.$ In the first round, the server transmits the following symbols, 
$ T_1: X^1_{2,3} \oplus X^4_{1,3} \oplus X^6_{1,2},$ $T_2: X^1_{2,4} \oplus X^4_{1,4} \oplus X^8_{1,2},$ $T_3: X^1_{3,4} \oplus X^6_{1,4} \oplus X^8_{1,3}$ and $T_4: X^4_{3,4} \oplus X^6_{2,4} \oplus X^8_{2,3}$. In the second round the transmissions are: $T_5: X^2_{2,3} \oplus X^5_{1,5} \oplus X^7_{1,2},$  $T_6: X^2_{2,4} \oplus X^5_{1,4},$ $T_7: X^2_{3,4} \oplus X^7_{1,4}$ and $T_8: X^5_{3,4} \oplus X^7_{2,4}$. The transmissions in the third round are: $T_9: X^3_{2,3},$ $T_{10}: X^3_{2,4}$ and $T_{11}: X^3_{3,4}$.  If we need to correct $\delta=1$ error, we need to concatenate SC transmission scheme with a classical error correcting code with optimal length. From \cite{Gra}, we have $N_2[11,3]=15.$ Hence the optimal concatenation can be done with a $[15,11,3]_2$ code. 
\end{exmp} 
\section{Improvement on SC Scheme for Non-distinct Demands}
\label{sec:new_scheme}
In this section, we consider the case when the demands are non-distinct. We give a delivery scheme which clearly has an advantage over the scheme in \cite{PUE}. We give an expression for the achievable rate for any demand vector $\mathbf{d}$ which meets the expression for achievable rate in the case of \cite{PUE} for the worst case demand. Before formally describing the proposed delivery scheme, we demonstrate the main ideas of the scheme through a motivating example.
\subsection{Motivating Example}
Consider the same system which we explained in Example \ref{Ex:alphasc}. The placement scheme and user assignments are the same as in Example \ref{Ex:alphasc}. We assume here that the demand vector $\mathbf{d}=(1,2,3,1,1,1,1,1)$. Thus, here $N_e(\mathbf{d})=3$. Before the delivery scheme starts, we eliminate some demands which are redundant. If multiple users which are connected to the same cache demand the same file, the delivery scheme need to satisfy the demand of one of them and the others also get what they want. Hence we can eliminate the repeated demand among the users which are connected to the same cache. Thus in the example, we can modify the association profile as $\mathcal{L}= (3,1,1,1)$ and $\mathbf{d}=(1,2,3,1,1,1)$. After this, the delivery scheme is done in rounds as in \cite{PUE}, but with a modification. Delivery takes place in 3 rounds, with each round respectively serving the following sets of users:
$\mathcal{R}_1=\{1,1,1,1\}, \mathcal{R}_2=\{2\}$ and $\mathcal{R}_3=\{3\}$. 
In the first round, the server transmits
$$ T_{\{1,1,1\}_1}= X^1_{2,3} \oplus X^1_{1,3} \oplus X^1_{1,2}$$ 
$$ T_{\{1,1,1\}_2}= X^1_{2,4} \oplus X^1_{1,4} \oplus X^1_{1,2}$$
$$ T_{\{1,1,1\}_3}= X^1_{3,4} \oplus X^1_{1,4} \oplus X^1_{1,3}.$$ 
Here the decoding is done as in \cite{YMA}. For instance, user 1, upon receiving $T_{\{1,1,1\}_1}$, can decode $X^1_{2,3}$ using the helper cache contents $X^1_{1,3}$ and $X^1_{1,2}$. Similarly using other transmissions, other subfiles can be decoded. In the second round, we have the following set of transmissions: 
$$ T_{2_1}= X^2_{2,3}$$
$$  T_{2_2}= X^2_{2,4} $$
$$ T_{2_3}= X^2_{3,4}. $$
In the last round the server serves user 3 with three more transmissions given by:
$$  T_{3_1}= X^3_{2,3} $$
$$  T_{3_2}= X^3_{2,4} $$
$$ T_{3_3}= X^3_{3,4}. $$
Hence there are a total of 9 transmissions, which means that the rate achieved is $\frac{9}{6}=\frac{3}{2}$. This is a smaller rate compared to the rate $\frac{11}{6}$ achieved by the scheme in \cite{PUE}. 
\subsection{General Delivery Phase}
\label{subsec:gen_sc_scheme}
We follow the assumptions and most of the notations as in \cite{PUE} to describe the scheme. 
Let the demand vector be $\mathbf{d}$ and let the number of distinct files requested be $N_e(\mathbf{d})$. We use the notation $N_e(\mathcal{U}_{\lambda})$ for the number of distinct files demanded by the users in $\mathcal{U}_{\lambda}$.  We need to consider only $N_e(\mathcal{U}_{\lambda})$ users which request distinct files and satisfy their demand. This is because, any other user in $\mathcal{U}_{\lambda}$ can get its requested file from the transmissions. Hence before the delivery starts, we eliminate the users with repeated demand from each $\mathcal{U}_{\lambda}$. After eliminating such users, let the modified association profile be $\mathbf{\mathcal{L}}'$. The remaining users associated to cache $\lambda$ is denoted by  $\mathcal{U}_{\lambda}'$. Moreover,  let $\mathcal{L'}_\lambda \triangleq|\mathcal{U}_{\lambda}'|$. Without loss of generality, we assume that  $\mathcal{L}'_1\geq \mathcal{L}'_2 \geq \ldots \mathcal{L}'_\Lambda.$ 
Delivery phase consists of $\mathcal{L}'_1$ rounds, where each round $j \in [\mathcal{L'}_1]$ serves users 
$$ \mathcal{R}'_j= \bigcup_{\lambda \in [\Lambda]} (\mathcal{U}'_\lambda (j): \mathcal{L'}_\lambda \geq j), $$
where $\mathcal{U}'_\lambda (j)$ is the $j$th user in the set $\mathcal{U}'_\lambda.$ Let the number of distinct files in $\mathcal{R}'_j$ be $N_e(\mathcal{R}'_j).$ For each round $j$, the server selects a subset of $N_e(\mathcal{R}'_j)$ users, denoted by $\mathcal{P}_j$ that requests $N_e(\mathcal{R}'_j)$ different files. These users are considered as leaders. For each round $j$, we create sets $\mathcal{Q} \subseteq [\Lambda]$ of size $|\mathcal{Q}| = \Lambda \gamma +1$, and for each set $\mathcal{Q}$ which satisfy $\mathcal{A} \cap \mathcal{P}_j \neq \phi,$ we pick the set of receiving users as
$$ \mathcal{E}_\mathcal{Q} = \bigcup_{\lambda \in \mathcal{Q}} (\mathcal{U}'_\lambda (j): \mathcal{L'}_\lambda \geq j). $$
If  $\mathcal{E}_\mathcal{Q} \neq \phi$, the server transmits,
$$ T_{\mathcal{E}_\mathcal{Q}} = \oplus_{\lambda \in \mathcal{Q}: {\mathcal{L'}_\lambda} \geq j} X^{d_{\mathcal{U}'_\lambda(j)}}_{\mathcal{Q}\setminus \{\lambda\}}. $$
If $\mathcal{E}_\mathcal{Q} = \phi$, there is no transmission. 
Since this transmission scheme uses scheme in \cite{YMA} for each round, the decoding at each receiver is ensured. 
The theorem below gives an expression for rate in this scheme.
\begin{thm}
\label{thm:rate_nond}
For coded caching problems with SC prefetching scheme,
\begin{align*}
R&(\mathcal{M}_{\text{SC}}, \delta=0) = \\ &Conv\bigg(\mathbb{E}_{\mathbf{d}}\bigg[ \frac{\sum_{j=1}^{\mathcal{L}_1} {\Lambda \choose \Lambda \gamma+1}-{\Lambda-N_e(\mathcal{R}'_j) \choose \Lambda \gamma+1}-{\Lambda-|{R}'_j| \choose \Lambda \gamma+1}}{{\Lambda \choose \Lambda \gamma}} \bigg]\bigg)
\end{align*}
at points $\gamma =\{\frac{1}{\Lambda}, \frac{2}{\Lambda}, \ldots, 1\}.$
\end{thm}
\begin{IEEEproof}
Since, $|\mathcal{Q}|=\Lambda \gamma +1$, there can be a total of ${\Lambda \choose \Lambda \gamma+1}$ sets of users $\mathcal{E}_{\mathcal{Q}}$. Furthermore, we see that there are ${\Lambda-|{R}'_j| \choose \Lambda \gamma+1}$ such sets that are empty. Moreover, since the transmissions are done only for such sets $\mathcal{Q}$ which satisfy $\mathcal{A} \cap \mathcal{P}_j \neq \phi,$ we see that each round $j$ consists of
$$  {\Lambda \choose \Lambda \gamma+1}-{\Lambda-N_e(\mathcal{R}'_j) \choose \Lambda \gamma+1}-{\Lambda-|{R}'_j| \choose \Lambda \gamma+1} $$ 
transmissions. Since each file is split into ${\Lambda \choose \Lambda \gamma}$ subfiles, the statement of the theorem follows.
\end{IEEEproof}
\subsection{Generalized Independence Number}
In this subsection, we find a bound for the generalized independence number $\alpha(\mathcal{M}_{\text{SC}}, \mathbf{d})$ of the index coding problems $\mathcal{I}(\mathcal{M}_{\text{SC}}, \mathbf{d})$, which covers even the case of non-distinct demands. From the rate expression in Theorem \ref{thm:rate_nond}, we have the upper bound for $\kappa(\mathcal{M}_{\text{SC}}, \mathbf{d})$ given by
\begin{align}
\begin{split}
\label{eq:kappa_general}
\kappa(\mathcal{M}_{\text{SC}}, &\mathbf{d})  \leq  \\ &\sum_{j=1}^{\mathcal{L}_1} {\Lambda \choose \Lambda \gamma+1}-{\Lambda-N_e(\mathcal{R}'_j) \choose \Lambda \gamma+1}-{\Lambda-|{R}'_j| \choose \Lambda \gamma+1}. 
\end{split}
\end{align}
The theorem below gives a lower bound for $\alpha(\mathcal{M}_{\text{SC}}, \mathbf{d})$.
\begin{thm}
For the index coding problems $\mathcal{I}(\mathcal{M}_{\text{SC}}, \mathbf{d})$,  
\begin{equation}
\label{eq:alpha_general}
 \alpha(\mathcal{M}_{\text{SC}}, \mathbf{d}) \geq \sum_{i=1}^{\Lambda-\Lambda \gamma} \mathcal{L}'_i {\Lambda-i \choose \Lambda \gamma}.
\end{equation}
\end{thm}
\begin{IEEEproof}
The modified association profile is $\mathcal{L'} =(\mathcal{L'}_1, \ldots, \mathcal{L'}_{\Lambda})$. Hence the  theorem follows from Theorem \ref{thm:alphakappascw}. 
\end{IEEEproof}

Since the expressions in  \eqref{eq:kappa_general} and \eqref{eq:alpha_general} are different, the equality of $\alpha(\mathcal{M}_{\text{SC}}, \mathbf{d})$ and $\kappa(\mathcal{M}_{\text{SC}}, \mathbf{d})$ cannot be guaranteed in general. There are cases when these become equal. In that case, an optimal error correcting delivery scheme is obtained by concatenation of the delivery scheme proposed in Section \ref{subsec:gen_sc_scheme} and an optimal error correcting code. This is illustrated in detail in the following example.
\begin{exmp}
Consider a shared cache system with $N=K=9$, $M=\Lambda =3$. Hence the parameter $\gamma = \frac{M}{N}= \frac{1}{3}$. Consider a uniform association profile $\mathcal{L}=(3,3,3).$ Each file $X^n$ is divided into ${3 \choose 1}=3$ subfiles $X^n_1, X^n_2$ and  $X^n_3$.   
We know that if all the files are demanded, the number of transmissions required by SC delivery scheme is $9$ from \eqref{eq:kappa_sc}. We assume that only $6$ files are demanded and let the demand vector be $\mathbf{d}=(1,2,2,3,4,4,5,6,6)$. We use the delivery scheme proposed in Section  \ref{subsec:gen_sc_scheme} as follows. We need to remove the repeated demand from each of the caches. Thus the modified association profile will be $\mathcal{L'}=(2,2,2).$ The corresponding modified demand vector will be $\mathbf{d}=(1,2,3,4,5,6)$. There will be $\mathcal{L'}_1=2$ rounds of transmissions. The first round serves the users in $\mathcal{R'}_1=\{1,3,5\}$ and the second round serves the users in $\mathcal{R'}_2=\{2,4,6\}.$ The transmissions in the first round are
$$ T_{\{1,3\}} = X^1_2 \oplus X^3_1 $$
$$ T_{\{1,5\}} = X^1_3 \oplus X^5_1 $$
$$ T_{\{3,5\}} = X^3_3 \oplus X^5_2. $$
The transmissions in the second round are given as
$$ T_{\{2,4\}} = X^2_2 \oplus X^4_1 $$
$$ T_{\{2,6\}} = X^2_3 \oplus X^6_1 $$
$$ T_{\{4,6\}} = X^4_3 \oplus X^6_2. $$
Hence there are $6$ transmissions. Hence for the index coding problem $\mathcal{I}(\mathcal{M}_{\text{SC}}, \mathbf{d})$, we have $\kappa(\mathcal{M}_{\text{SC}}, \mathbf{d}) \leq 6.$

For finding a lower bound for $\alpha(\mathcal{M}_{\text{SC}}, \mathbf{d})$, we construct the set $B(\mathbf{d})$ as in the proof of Theorem \ref{thm:alphakappascw}. We obtain the set $B(\mathbf{d})$ as follows:
$$ B(\mathbf{d})=\{X^1_2, X^1_3, X^2_2, X^2_3, X^3_3, X^4_3 \}.  $$
From this, we get $\alpha(\mathcal{M}_{\text{SC}}, \mathbf{d}) \geq 6.$ Thus for this case, $\alpha$ and $\kappa$ bounds meet. Hence for this case, the optimal linear error correcting delivery scheme is to concatenate the improved scheme in Section \ref{subsec:gen_sc_scheme} with an optimal linear error correcting code. For instance, suppose that we want to correct $\delta =1$ transmission error. From \cite{Gra}, we get $N_2[6,3]=10.$ Hence the concatenation can be done with a $[10,6,3]_2$ linear code to obtain  an optimal linear error correcting delivery scheme.

Assume now that only 5 files are demanded and the demand vector is $\mathbf{d}= (1,2,3,4,5,1,2,3,4)$. Here since there is no repeated demand within a cache, there is no user to be eliminated. The transmission is carried out in 3 rounds. The users served in the three rounds are given by
$$\mathcal{R'}_1=\{1,1,2\}, $$
$$\mathcal{R'}_2=\{2,4,4\} \text{ and } $$
$$\mathcal{R'}_3=\{3,5,3\}. $$
The transmissions in each round is done according to the improved scheme. The transmissions in first round are:
$$ T_{\{1,1\}} = X^1_2 \oplus X^1_1, $$
$$ T_{\{1,2\}_1} = X^1_3 \oplus X^2_1 \text{ and } $$
$$ T_{\{1,2\}_2} = X^1_3 \oplus X^2_2. $$
The transmissions in the second round are:
$$ T_{\{2,4\}_1} = X^2_2 \oplus X^4_1, $$
$$ T_{\{2,4\}_2} = X^2_3 \oplus X^4_1 \text{ and } $$
$$ T_{\{4,4\}} = X^4_3 \oplus X^4_2. $$
The transmissions in the third round are:
$$ T_{\{3,5\}} = X^3_2 \oplus X^5_1, $$
$$ T_{\{3,3\}} = X^3_3 \oplus X^3_1 \text{ and } $$
$$ T_{\{5,3\}} = X^5_3 \oplus X^3_2. $$
Hence there are 9 transmissions. Thus for the index coding problem $\mathcal{I}(\mathcal{M}_{\text{SC}}, \mathbf{d})$, we have $\kappa(\mathcal{M}_{\text{SC}}, \mathbf{d}) \leq 9.$

The set $B(\mathbf{d})$ is constructed for this case as
$$ B(\mathbf{d})=\{X^1_2, X^1_3, X^2_2, X^2_3, X^3_2, X^3_3, X^4_3, X^5_3 \}.  $$
From this, we get that $\alpha(\mathcal{M}_{\text{SC}}, \mathbf{d}) \geq 8.$ Thus, for this case we cannot conclude that $\alpha$ and $\kappa$ bounds meet. Hence the concatenation scheme may not be optimal. 
\end{exmp} 

\section{Conclusion}
We considered the SC scheme and for worst case demand, we proved that for all the corresponding index coding problems, the $\alpha$ and $\kappa$ bounds meet. This makes the concatenation of SC delivery scheme with an optimal classical error correcting code which corrects the required number of errors to be optimal. Moreover, for the case of non-distinct demands, we proposed an improved scheme which has clear advantage over the scheme in \cite{PUE}. 

%%%%%%%%%%%%%%%%%%%%%%%%%%%%%%%%%%%%%%%%%%%%%%%%%%%%%%%%%%%%%%%%%%%%%

%%%%%%%%%%%%%%%%%%%%%%%%%%%%%%%%%%%%%%%%%%%%%%%%%%%%%%%%%%%%%%%%%%%%%
\section*{Acknowledgment}
This work was supported partly by the Science and Engineering Research Board (SERB) of Department of Science and Technology (DST), Government of India, through J.C. Bose National Fellowship to B. Sundar Rajan.

\end{document}